\documentclass{jpsj3}
\usepackage{txfonts}
\usepackage{color}

\title{Formulation of Entropy through Work by Carnot Machine and Direct Derivation of Law of Entropy Non-Decrease from Kelvin-Planck Principle}

\author{Yuki Izumida\thanks{izumida@k.u-tokyo.ac.jp}}
\inst{Department of Complexity Science and Engineering, Graduate School of Frontier Sciences, The University of Tokyo, Kashiwa 277-8561, Japan} 

\abst{We derive the law of entropy non-decrease directly from the Kelvin-Planck principle for simple and compound systems without using the Clausius inequality.
A key of the derivation is a new formulation of entropy in terms of work by a Carnot machine operating between a system and a single heat reservoir at fixed temperature, which is equivalent to Clausius entropy based on heat and Gyftopoulos-Beretta entropy based on work.
We also show that we may characterize entropy as an extra thermodynamic cost that needs to be paid to create nonuniformity in the system.}

\begin{document}
\maketitle

\section{Introduction}

How information of irreversibility is encoded into entropy as a state quantity?
This would be one of the most difficult points for those who learn thermodynamics first.
This is also true for physicists today, and the famous paper by Lieb and Yngvason was aimed to elucidate the nature of entropy 
in terms of adiabatic accessibility in their axiomatic approach to thermodynamics~\cite{LY1999,LY1998,LY2000}, where neither statistical mechanics nor heat engines are used~\cite{LY2000} (see also Ref.~\cite{T2011}).

Here, let us recall a conventional approach to entropy.
Consider a thermally isolated system in an equilibrium state $A_1$.  
After an arbitrary adiabatic process that changes the system's state to another equilibrium state $A_2$, 
the system's entropy $S$ never decreases before and after the change:
\begin{eqnarray}
S_2-S_1 \ge 0,\label{eq.S_non}
\end{eqnarray}
which is the law of entropy non-decrease, and the equality applies to reversible adiabatic processes.
Here, the entropies of the system's states $A_1$ and $A_2$ are given by
\begin{eqnarray}
S_1=S_0+\int_0^1 \frac{\delta Q}{T_R}, \ \ S_2=S_0+\int_0^2 \frac{\delta Q}{T_R},\label{eq.S_def}
\end{eqnarray}
respectively, where $S_0$ is the entropy at a reference state $A_0$, $\delta Q$ denotes heat supplied to the system, and $T_R$ denotes the temperature of a heat reservoir.
During the process from $A_0$ to $A_1$, the temperature $T_R$ of the heat reservoir that contacts the system may change.
Equation~(\ref{eq.S_def}) is the formula derived by Clausius.
Usually, in standard textbooks on thermodynamics such as Ref.~\cite{F1956}, after introducing Carnot cycle and Carnot efficiency as the maximum efficiency of heat engines, Eq.~(\ref{eq.S_non}) is derived by using the Clausius inequality for a cyclic process:
\begin{eqnarray}
\oint \frac{\delta Q}{T_R}\le 0.\label{eq.Clausius_ineq}
\end{eqnarray}
The existence of entropy as a state quantity (\ref{eq.S_def}) is shown by using the equality condition of the Clausius inequality.
Then, by applying the Clausius inequality Eq.~(\ref{eq.Clausius_ineq})
to a cyclic process $A_1 \to A_2 \stackrel{\rm rev}{\longrightarrow} A_1$, where ``{\rm rev}" implies that the process is reversible, we have
\begin{eqnarray}
S_2-S_1-\int_\Gamma \frac{\delta Q}{T_R}\ge 0.\label{eq.Clausius}
\end{eqnarray}
Here, $\Gamma$ denotes a path from $A_1$ to $A_2$.
If the process $A_1 \to A_2$ is an adiabatic process, which we write as $A_1 \stackrel{\rm ad}{\longrightarrow} A_2$, we may put $\delta Q=0$ into the above equality and derive Eq.~(\ref{eq.S_non}).

The sketch of the above derivation of the law of entropy non-decrease is based on the Clausius inequality Eq.~(\ref{eq.Clausius_ineq}).
The Clausius inequality is derived from the Kelvin-Planck principle, which manifests one expression of the second law of thermodynamics: 
We cannot extract a positive work from a cyclic process with a single heat reservoir at fixed temperature:
\begin{eqnarray}
W_{\rm cyc} \le 0,\label{eq.Kelvin}
\end{eqnarray}
where $W_{\rm cyc}$ denotes the work extracted from a cycle starting from an equilibrium state, and the equality applies to reversible cycles.
Deriving the Clausius inequality from the Kelvin-Planck principle relies on a complicated thermodynamic process 
that tactically involves multiple (or infinitely many) heat reservoirs at different temperatures and multiple (or infinitely many) Carnot heat engines~\cite{F1956}.
This is because the Kelvin-Planck principle is originally applied to a process with a single heat reservoir at fixed temperature, and a complicated setup is necessary for it to be applied to a case with multiple (or infinitely many) heat reservoirs.
Moreover, the law of entropy non-decrease Eq.~(\ref{eq.S_non}) is derived as a special case of the Clausius inequality Eq.~(\ref{eq.Clausius_ineq}) (i.e., the case of $\delta Q=0$ in Eq.~(\ref{eq.Clausius})) and it is thus derived {\it indirectly} from the Kelvin-Planck principle.

In this paper, we show that we can derive the law of entropy non-decrease Eq.~(\ref{eq.S_non}) directly from the Kelvin-Planck principle Eq.~(\ref{eq.Kelvin}) without using the Clausius inequality Eq.~(\ref{eq.Clausius_ineq}).
To this end, we give a new formulation of entropy of a substance in terms of work of a Carnot machine operating between the substance and a single heat reservoir at fixed temperature.
Because we will assume only a single heat reservoir at fixed temperature in our derivation, it apparently makes it possible to apply the Kelvin-Planck principle to show the law of entropy non-decrease. 
We discuss the relationship between our entropy and the Clausius entropy (\ref{eq.S_def}) based on heat or the Gyftopoulos-Beretta entropy~\cite{GB2005} based on work. 
Moreover, we also show that entropy may be characterized by an extra thermodynamic cost that needs to be paid to create nonuniformity in the system, 
which makes it possible to distinguish between two macroscopic systems of the same internal energy and volume.

\section{Main results}\label{sec_derivation}

We derive the law of entropy non-decrease Eq.~(\ref{eq.S_non}) from the Kelvin-Planck principle Eq.~(\ref{eq.Kelvin}).
After the formulation of our entropy in terms of work by a Carnot machine in Sec.~\ref{sec_entropy}, we give the derivation for simple systems in Sec~\ref{sec_simple} and compound systems in Sec.~\ref{sec_compound}.
Here, a simple system is a system in an equilibrium state that can be specified with the system's temperature $T$ and volume $V$.
A compound system is a system composed of simple systems.

\subsection{Formulation of entropy as state quantity}\label{sec_entropy}
\begin{figure}[t!]
\centering
\includegraphics[scale=0.55]{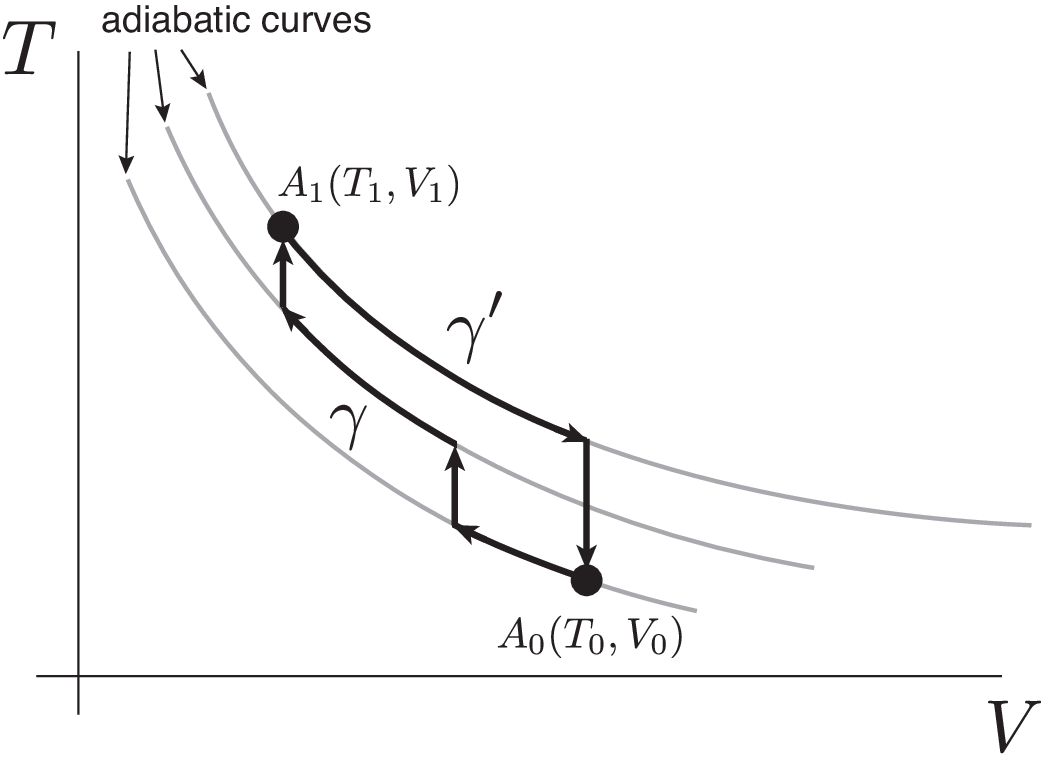}
\caption{A reversible path $\gamma$ from $A_0$ to $A_1$ and a reversible path $\gamma'$ from $A_1$ to $A_0$ on the temperature-volume ($T$--$V$) plane. They consist of the reversible adiabatic processes (arrows along the adiabatic curves) and the reversible temperature-changing processes with a Carnot machine operating between the system with its volume being kept fixed and a heat reservoir at temperature $T_o$ (vertical arrows). 
}
\label{fig_entropy}
\end{figure}

We introduce our entropy as a state quantity in the following way.
Consider a simple system initially in an equilibrium state $A_0$ specified with $(T_0, V_0)$ with $T_0$ and $V_0$ the system's initial temperature and volume, respectively.
Consider a reversible path $\gamma$ from the equilibrium state $A_0$ to another equilibrium state $A_1$ specified with $(T_1, V_1)$ 
and a reversible path $\gamma'$ from the state $A_1$ to $A_0$ (Fig.~\ref{fig_entropy}). 
Any reversible path $\gamma$ from the equilibrium state $A_0$ to another one $A_1$ can be realized by a combination of 
(i) a reversible adiabatic process (the arrow along the adiabatic curve in Fig.~\ref{fig_entropy}) and (ii) a reversible temperature-changing process with a Carnot machine operating between the system with its volume being kept constant and a single heat reservoir at {\it fixed} temperature $T_o$ to raise or lower the system's temperature (the vertical arrow in Fig.~\ref{fig_entropy})
by regarding the system as a finite-sized heat reservoir~\cite{note1}; any reversible path $\gamma$ may be decomposed into $N$-reversible processes $\gamma=\sum_{i=0}^{N-1} \gamma_i$ with $\gamma_i$ being the $i$-th reversible process, which is either of the above two thermodynamic processes (i) and (ii). 
In particular, when the path is continuous, it may be realized by a combination of the processes (i) and (ii) by taking the limit of $N \to \infty$.
It should be noted that there is no need for the system's temperature to match the fixed temperature $T_o$ of the heat reservoir during the process including the initial and final states $A_0$ and $A_1$.

The reversible works $W_{0\to 1}^{\rm rev}$ and $W_{1\to 0}^{\rm rev}$ extracted from $\gamma$ and $\gamma'$, respectively, are given by
\begin{eqnarray}
W_{0\to 1}^{\rm rev}[\gamma]=U_0-U_1+T_o\int_\gamma \frac{\delta Q}{T},\label{eq.W_gamma}
\end{eqnarray}
and
\begin{eqnarray}
W_{1\to 0}^{\rm rev}[\gamma']=U_1-U_0+T_o\int_{\gamma'} \frac{\delta Q}{T},\label{eq.W_gamma_dash}
\end{eqnarray}
where $U$ is the internal energy of the system and $T$ is the temperature of the system, {\it not} the temperature of the heat reservoir; the heat reservoir at fixed temperature $T_o$ is used here. We will derive Eqs.~(\ref{eq.W_gamma}) and (\ref{eq.W_gamma_dash}) later.

By applying the equality case of the Kelvin-Planck principle Eq.~(\ref{eq.Kelvin}) to the reversible cycle $A_0 \stackrel{\rm rev}{\longrightarrow} A_1 \stackrel{\rm rev}{\longrightarrow} A_0$, we have
\begin{eqnarray}
W_{\rm cyc}=W_{0\to 1}^{\rm rev}[\gamma]+W_{1\to 0}^{\rm rev}[\gamma']=0.\label{eq.Kelvin_rev}
\end{eqnarray}
From Eqs.~(\ref{eq.W_gamma})--(\ref{eq.Kelvin_rev}), we have
\begin{eqnarray}
\int_\gamma \frac{\delta Q}{T}=-\int_{\gamma'} \frac{\delta Q}{T}=\int_{-\gamma'} \frac{\delta Q}{T},\label{eq.Q_T}
\end{eqnarray}
where $-\gamma'$ denotes the reversed path of $\gamma'$.
Because $\gamma$ and $-\gamma'$ are arbitrary paths connecting $A_0$ and $A_1$, Eq.~(\ref{eq.Q_T}) implies that the quantity $\int_\gamma \delta Q/T$ has a unique value independent of a path, which we denote by $\int_0^1 \delta Q/T$.
Therefore, we can define a state quantity $S$, which takes $S_1$ at the state $A_1$ as
\begin{eqnarray}
S_1=S_0+\int_0^1 \frac{\delta Q}{T}\label{eq.S_Kelvin}
\end{eqnarray}
with $S_0$ at the state $A_0$ a criterion value. This state quantity $S$ is our entropy. 
We can write the reversible work $W_{0\to 1}^{\rm rev}$ from $A_0$ to $A_1$ in Eq.~(\ref{eq.W_gamma}) as
\begin{eqnarray}
W_{0\to 1}^{\rm rev}=U_0-U_1-T_o(S_0-S_1)\label{eq.W_gamma_S}
\end{eqnarray}
in a form independent of a path $\gamma$ by using the entropy. 
Equation (\ref{eq.W_gamma_S}) serves as the maximum work extracted from a system with the aid of a heat reservoir at fixed temperature (see Appendix~\ref{sec_appendix} for a simple proof based on Ref.~\cite{LY2013}).
It should be noted again that the temperature $T$ in Eq.~(\ref{eq.S_Kelvin}) is the system's temperature, not the heat reservoir's one such as in Eq.~(\ref{eq.S_def}).
While Eq.~(\ref{eq.S_def}) is usually obtained via the equality case of the Clausius inequality Eq.~(\ref{eq.Clausius_ineq}) that assumes thermal contact with multiple (or infinitely many) heat reservoirs with different temperatures, Eq.~(\ref{eq.S_Kelvin}) has been obtained by directly applying the equality case of the Kelvin-Planck principle assuming only a single heat reservoir at fixed temperature $T_o$.
But, once Eq.~(\ref{eq.S_Kelvin}) is established, we may interpret that it is equivalent to the Clausius entropy Eq.~(\ref{eq.S_def}), 
as any path in Fig.~\ref{fig_entropy} can be realized with a reversible process during which a system contacts multiple (or infinitely many) heat reservoirs with different temperatures, and the system's temperature $T$ agrees with the temperature $T_R$ of each heat reservoir for such a reversible process.

Now, we derive Eqs.~(\ref{eq.W_gamma}) and (\ref{eq.W_gamma_dash}).
The reversible work $\Delta W_i^{\rm rev}[\gamma_i]$ extracted from the $i$-th reversible process $\gamma_i$ from $(T^{(i)}, V^{(i)})$ to $(T^{(i+1)}, V^{(i+1)})$ as the part of $\gamma=\sum_{i=0}^{N-1} \gamma_i$, where $T^{(0)}=T_0$ and $T^{(N)}=T_1$, is given by the following two cases:
\begin{eqnarray}
({\rm i}) \ \Delta W_i^{\rm rev}[\gamma_i]&&=U(T^{(i)}, V^{(i)})-U(T^{(i+1)}, V^{(i+1)}),\label{eq.DW_adi}\\
({\rm ii}) \ \Delta W_i^{\rm rev}[\gamma_i]&&=U(T^{(i)}, V^{(i)})-U(T^{(i+1)}, V^{(i+1)})+T_o\int_{\gamma_i}\frac{\delta Q}{T},\label{eq.DW_Car}
\end{eqnarray}
depending on the type of the $i$-th process. Here, $\Delta W_i^{\rm rev}[\gamma_i]$ in Eq.~(\ref{eq.DW_adi}) for the process (i) is independent of $\gamma_i$ (i.e., determined only by the initial and final system's states), and note that $U(T^{(i+1)}, V^{(i+1)})=U(T^{(i+1)}, V^{(i)})$ in Eq.~(\ref{eq.DW_Car}) because of $V^{(i+1)}=V^{(i)}$ (fixed volume) for the process (ii).
These works (\ref{eq.DW_adi}) and (\ref{eq.DW_Car}) are obtained as follows: 
The work $\Delta W_i^{\rm rev}[\gamma_i]$ in Eq.~(\ref{eq.DW_adi}) for the process (i) is just given by the internal energy change $U^{(i)}-U^{(i+1)}=U(T^{(i)},V^{(i)})-U(T^{(i+1)},V^{(i+1)})$.
For the work $\Delta W_i^{\rm rev}[\gamma_i]$ in Eq.~(\ref{eq.DW_Car}) for the process (ii),
we operate a Carnot machine (heat engine, refrigerator, or heat pump) between the system at temperature $T$ and the heat reservoir at temperature $T_o$ until the system's temperature $T$ initiating from $T^{(i)}$ reaches $T^{(i+1)}$ by keeping $V=V^{(i)}$.
To raise the system's temperature $T$, we may operate a Carnot heat engine for $T<T_o$ and a Carnot heat pump for $T_o \le T$; to lower the system's temperature $T$, we may operate a Carnot heat engine for $T_o<T$ and a Carnot refrigerator for $T\le T_o$.
For any case, the infinitesimal work $\delta W$ extracted by a Carnot machine operating between the system at $T$ and the heat reservoir at $T_o$ is commonly given by
\begin{eqnarray}
\delta W=-\left(1-\frac{T_o}{T}\right)\delta Q,\label{eq.delta_W}
\end{eqnarray}
where $\delta W$ and $\delta Q$ take a positive or negative value depending on the situation.
Then, the reversible work $\Delta W_i^{\rm rev}[\gamma_i]$ in Eq.~(\ref{eq.DW_Car}) extracted from the process (ii) in the total is given by the integration of Eq.~(\ref{eq.delta_W}):
\begin{eqnarray}
\Delta W_i^{\rm rev}[\gamma_i]=\int_{\gamma_i}\delta W&&=-\int_{\gamma_i} \left(1-\frac{T_o}{T}\right)\delta Q\nonumber\\
&&=-\int_{\gamma_i}\delta Q+T_o\int_{\gamma_i} \frac{\delta Q}{T}\nonumber\\
&&=U(T^{(i)},V^{(i)})-U(T^{(i+1)}, V^{(i+1)})+T_o \int_{\gamma_i} \frac{\delta Q}{T}.
\label{eq.W_0dash_1}
\end{eqnarray}
Here, in the last equality, we have used the fact that the total heat supplied to the system is equal to the internal energy change for the case of the vanishing work due to the fixed volume $V^{(i+1)}=V^{(i)}$.

By summing the work $\Delta W_i^{\rm rev}[\gamma_i]$ of each process (Eqs.~(\ref{eq.DW_adi}) and (\ref{eq.DW_Car})) along a given path $\gamma$ from $A_0$ to $A_1$, we have
\begin{eqnarray}
W_{0\to 1}^{\rm rev}[\gamma]&&=\sum_{i=0}^{N-1}\Delta W_i^{\rm rev}[\gamma_i] \nonumber\\
&&=\sum_{i=0}^{N-1}U(T^{(i)},V^{(i)})-U(T^{(i+1)}, V^{(i+1)})+T_o \int_{\sum_{i=0}^{N-1}\gamma_i}\frac{\delta Q}{T} \nonumber\\
&&=U_0-U_1+T_o \int_\gamma \frac{\delta Q}{T},\label{eq.W01}
\end{eqnarray}
which is Eq.~(\ref{eq.W_gamma}).  Equation~(\ref{eq.W_gamma_dash}) can also be calculated in the same manner.

\subsection{Derivation for simple systems}\label{sec_simple}

\begin{figure}[t!]
\centering
\includegraphics[scale=0.55]{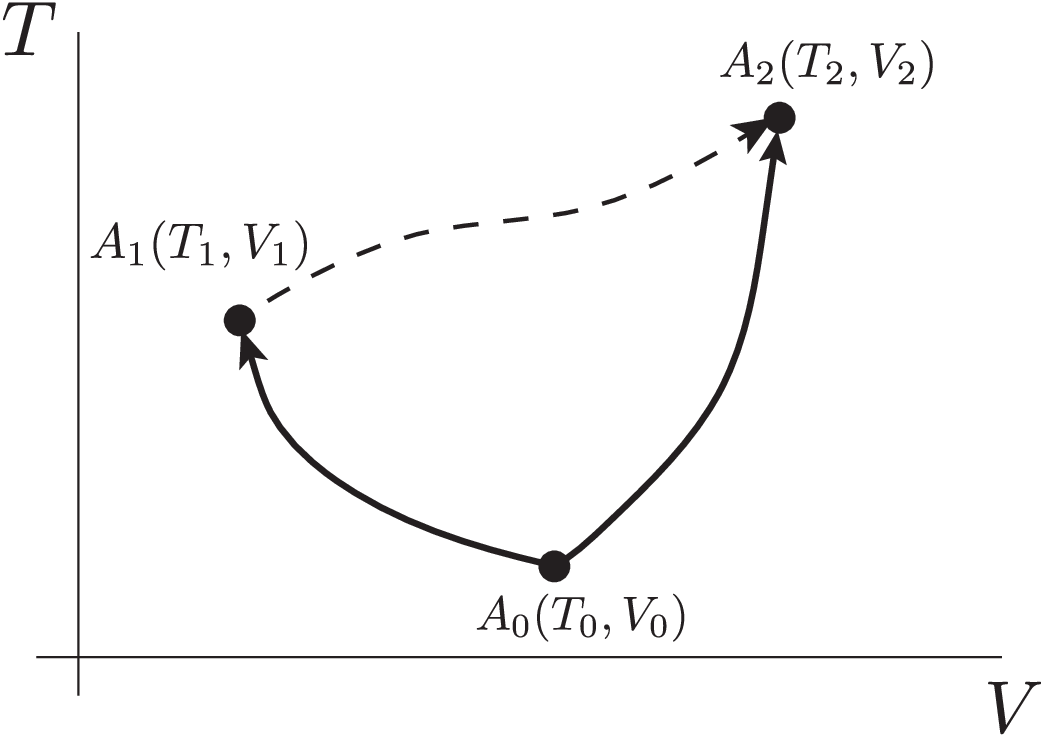}
\caption{An arbitrary adiabatic process from one equilibrium state $A_1$ of a simple system to another equilibrium state $A_2$ (dashed curve), where $A_1$ and $A_2$ are created reversibly from a reference equilibrium state $A_0$ (solid curves).
The Kelvin-Planck principle Eq.~(\ref{eq.Kelvin}) is applied to the cyclic process $A_0  \stackrel{\rm rev}{\longrightarrow}  A_1  \stackrel{\rm ad}{\longrightarrow} A_2 \stackrel{\rm rev}{\longrightarrow} A_0$.
}
\label{fig_simple}
\end{figure}

Now that we have formulated our entropy in Eq.~(\ref{eq.S_Kelvin}), we are ready to show its non-decrease during adiabatic processes. 
That is, we show Eq.~(\ref{eq.S_non}) from Eq.~(\ref{eq.Kelvin}). Here, we show it for simple systems.

Let a thermally isolated system be in the equilibrium state $A_1$ specified with $(T_1, V_1)$. 
Consider an arbitrary adiabatic process from $A_1$ to another equilibrium state $A_2$ with $(T_2, V_2)$ (Fig.~\ref{fig_simple}) by adiabatic operations such as stirring and moving a piston attached to the system, during which the system may be in a nonequilibrium state; after the operations, we may wait for a sufficiently long time until the system reaches the equilibrium state $A_2$.
The work extracted from this process $W_{1\to 2}^{\rm ad}$ is given by
\begin{eqnarray}
W_{1\to 2}^{\rm ad}=U_1-U_2,\label{eq.W12}
\end{eqnarray}
irrespective of whether the adiabatic process $A_1 \stackrel{\rm ad}{\longrightarrow} A_2$ is reversible or irreversible.

Let $A_0$ be a reference equilibrium state of the system specified with $(T_0, V_0)$.
We {\it create} the equilibrium state $A_1$ from the reference equilibrium state $A_0$ reversibly by combining two thermodynamic processes (i) and (ii) with the aid of a single heat reservoir at fixed temperature $T_o$ considered in Sec.~\ref{sec_entropy} (Fig.~\ref{fig_simple}).
The work extracted from this process is given by Eq.~(\ref{eq.W_gamma_S}):
\begin{eqnarray}
W_{0\to 1}^{\rm rev}=U_0-U_1-T_o(S_0-S_1).\label{eq.W01_2}
\end{eqnarray}

Subsequently, we consider a reversible thermodynamic process from $A_2$ to $A_0$ (Fig.~\ref{fig_simple}) and the reversible work $W_{2\to 0}^{\rm rev}$ extracted from it.
Because this process is reversible, we may consider the process from $A_0$ to $A_2$ and reverse the sign of $W_{0\to 2}^{\rm rev}$. 
In the similar manner to $W_{0\to 1}^{\rm rev}$, we obtain
\begin{eqnarray}
W_{2\to 0}^{\rm rev}=-W_{0\to 2}^{\rm rev}=-(U_0-U_2-T_o(S_0-S_2)).\label{eq.W02}
\end{eqnarray}

From Eqs.~(\ref{eq.W12}), (\ref{eq.W01_2}), and (\ref{eq.W02}), we obtain the work extracted from the whole cyclic process $A_0 \stackrel{\rm rev}{\longrightarrow} A_1 \stackrel{\rm ad}{\longrightarrow} A_2 \stackrel{\rm rev}{\longrightarrow} A_0$ as
\begin{eqnarray}
W_{\rm cyc}=W_{0\to 1}^{\rm rev}+W_{1\to 2}^{\rm ad}+W_{2\to 0}^{\rm rev}=T_o(S_1-S_2).\label{eq.Wcyc}
\end{eqnarray}
By applying the Kelvin-Planck principle Eq.~(\ref{eq.Kelvin}) to Eq.~(\ref{eq.Wcyc}), we derive the law of entropy non-decrease Eq.~(\ref{eq.S_non}).

\subsection{Derivation for compound systems}\label{sec_compound}

We show Eq.~(\ref{eq.S_non}) from Eq.~(\ref{eq.Kelvin}) for a compound system. 
Here, a compound system is composed of $M$ subsystems in equilibrium with different temperatures,
where each subsystem is separated by other subsystems by insulated walls.
So, a compound system may be regarded a nonuniform system.
The state of $j$-th subsystem is specified with $(T_j, V_j)$ ($j=1, \cdots, M$).
The internal energy $U$, entropy $S$, and volume $V$ of the compound system are given by the sum of those of subsystems as $U=\sum_{j=1}^M U_j$, and $S=\sum_{j=1}^M S_j$, and $V=\sum_{j=1}^M V_j$, respectively.

In the same manner as the simple system, we consider a cyclic process that consists of three thermodynamic processes $A_1 \stackrel{\rm ad}{\longrightarrow} A_2$, $A_0 \stackrel{\rm rev}{\longrightarrow} A_1$, and $A_2 \stackrel{\rm rev}{\longrightarrow} A_0$ as follows. 
Let $A_1$ be the state of a compound system specified with the states of $M_1$ subsystems $\{(T_{1, 1}, V_{1, 1}), (T_{1, 2}, V_{1, 2}), \cdots, (T_{1, M_1}, V_{1, M_1})\}$ in equilibrium, where $V_1=\sum_{j=1}^{M_1}V_{1, j}$ is the volume of the compound system at $A_1$.
Consider an arbitrary adiabatic process $A_1 \stackrel{\rm ad}{\longrightarrow} A_2$, where $A_2$ is specified with $M_2$ subsystems $\{(T_{2, 1}, V_{2, 1}), (T_{2, 2}, V_{2, 2}), \cdots, (T_{2, M_2}, V_{2, M_2})\}$ in equilibrium.
During this process, in addition to the adiabatic operations such as stirring, moving a piston attached to the system, 
and moving insulated walls partitioning subsystems, 
the number of subsystems is allowed to change from $M_1$ to $M_2$ by newly inserting or removing walls partitioning subsystems without any work cost.
The work extracted from this process $W_{1\to 2}^{\rm ad}$ is given by
\begin{eqnarray}
W_{1\to 2}^{\rm ad}=U_1-U_2,\label{eq.W12_cmp}
\end{eqnarray}
where $U_1=\sum_{j=1}^{M_1}U_{1, j}$ and $U_2=\sum_{j=1}^{M_2}U_{2, j}$.

Subsequently, we create the state $A_1$ of the compound system from a reference equilibrium state $A_0$ of a {\it simple} system specified with $(T_0, V_0)$.
First, we change the volume from $V_0$ to $V_1$ by a reversible adiabatic process, resulting in the temperature $T_{0'}$, where we denote the state $(T_{0'}, V_1)$ by $A_{0'}$.
The reversible work extracted from this process $A_0 \stackrel{\rm rev}{\longrightarrow} A_{0'}$ is given by
\begin{eqnarray}
W_{0\to0'}^{\rm rev}=U_0-U_{0'}.\label{eq.W_0_0dash_com}
\end{eqnarray}
Then, we insert insulated walls (without any work cost) so that the system is divided into $M_1$ subsystems, where each subsystem is in the equilibrium state $A_{0', j}$ specified with $(T_{0'}, V_{1, j})$ ($j=1, \dots, M_1$). 
By operating a Carnot machine between each subsystem and the heat reservoir at temperature $T_o$ until it reaches the equilibrium state with $(T_{1,j}, V_{1,j})$, 
the total reversible work extracted from this process $A_{0'} \stackrel{\rm rev}{\longrightarrow} A_1$ is given by
\begin{eqnarray}
W_{0'\to 1}^{\rm rev}&&=\sum_{j=1}^{M_1}(U_{0',j}-U_{1,j}-T_o(S_{0',j}-S_{1,j}))\nonumber\\
&&=U_{0'}-U_1-T_o(S_{0'}-S_1).\label{eq.W_idash_1_com}
\end{eqnarray}
By summing Eqs.~(\ref{eq.W_0_0dash_com}) and (\ref{eq.W_idash_1_com}), we obtain
\begin{eqnarray}
W_{0\to 1}^{\rm rev}=U_0-U_1-T_o(S_{0'}-S_1).\label{eq.W01_com}
\end{eqnarray}

Finally, the reversible work extracted from $A_2 \stackrel{\rm rev}{\longrightarrow} A_0$ should be calculated. 
Here, the system at the state $A_2$ consists of $M_2$ subsystems in equilibrium, 
whose state we denote by $A_{2,j}$ specified with $(T_{2, j}, V_{2, j})$, where $\sum_{j=1}^{M_2}V_j=V_2$. 
In the similar manner to the calculation of $W_{0\to 1}^{\rm rev}$ in Eq.~(\ref{eq.W01_com}),
we can calculate the reversible work extracted from $A_2 \stackrel{\rm rev}{\longrightarrow} A_0$ as the minus of $W_{0\to 2}^{\rm rev}$:
\begin{eqnarray}
W_{2\to 0}^{\rm rev}=-W_{0\to 2}^{\rm rev}=-(U_0-U_2-T_o(S_{0''}-S_2)),\label{eq.W02_com}
\end{eqnarray}
where $S_{0''}$ is the entropy at the state $A_{0''}$ specified with $(T_{0''}, V_2)$ reached from $A_0$ in a reversible adiabatic process.
Note that $S_{0'}=S_{0''}$ as $A_{0'}$ and $A_{0''}$ (and $A_0$) are on the same adiabatic curve.
By summing Eqs.~(\ref{eq.W12_cmp}), (\ref{eq.W01_com}), and (\ref{eq.W02_com}), and using $S_{0'}=S_{0''}$, we obtain the work extracted from the whole cyclic process $A_0 \stackrel{\rm rev}{\longrightarrow} A_1 \stackrel{\rm ad}{\longrightarrow} A_2 \stackrel{\rm rev}{\longrightarrow} A_0$ as
\begin{eqnarray}
W_{\rm cyc}=W_{0\to 1}^{\rm rev}+W_{1\to 2}^{\rm ad}+W_{2\to 0}^{\rm rev}=T_o(S_1-S_2).\label{eq.Wcyc_com}
\end{eqnarray}
By applying the Kelvin-Planck principle in Eq.~(\ref{eq.Kelvin}) to Eq.~(\ref{eq.Wcyc_com}), we derive the law of entropy non-decrease in Eq.~(\ref{eq.S_non}) for the compound system.

\section{Discussion}
\subsection{Gyftopoulos-Beretta entropy}
It should be noted that our formulation of entropy and its non-decrease in Sec.~\ref{sec_derivation} may share some similarities with the approach by Gyftopoulos and Beretta (GB)~\cite{GB2005} (see also Ref.~\cite{HG1976}).
By using notations in our formulation, their entropy is {\it defined} through the difference between works extracted from processes with and without the aid of a heat reservoir at fixed temperature $T_o$~\cite{note2}:
\begin{eqnarray}
&&S_1=S_0+\frac{W_{0\to 1}^{\rm rev}-(U_0-U_1)}{T_o},\ S_2=S_0+\frac{W_{0\to 2}^{\rm rev}-(U_0-U_2)}{T_o}.\label{eq.GB_S}
\end{eqnarray}
where we used $W_{0\to 1}^{\rm ad}=U_0-U_1$ and $W_{0\to 2}^{\rm ad}=U_0-U_2$ for the process without the aid of a heat reservoir.
This definition is equivalent to Eq.~(\ref{eq.S_Kelvin}) and is derived from it with the use of Eq.~(\ref{eq.W_gamma}),
 but the one focusing on work rather than heat.
This is a natural definition as Eq.~(\ref{eq.W_gamma_S}) suggests that the entropy serves to determine the reversible work as the {\it maximum} work extracted from a system with the aid of a heat reservoir at temperature $T_o$~\cite{LY2013}.
Moreover, it should be noted that the GB entropy is referred to as the same entropy as the one derived by Lieb and Yngvason in their axiomatic thermodynamics~\cite{LY2013} (see also Ref.~\cite{ZB2014} for a related discussion).

By taking the difference between the two in Eq.~(\ref{eq.GB_S}), we get
\begin{eqnarray}
S_2-S_1
=\frac{W_{0\to 2}^{\rm rev}-W_{0\to 1}^{\rm rev}-(U_1-U_2)}{T_o}.\label{eq.GB_Snd}
\end{eqnarray}
Applying the Kelvin-Planck principle to a cyclic process $A_0 \stackrel{\rm rev}{\longrightarrow} A_1 \stackrel{\rm ad}{\longrightarrow} A_2 \stackrel{\rm rev}{\longrightarrow} A_0$, we have
\begin{eqnarray}
&&W_{0\to 1}^{\rm rev}+(U_1-U_2)+W_{2\to 0}^{\rm rev}\le 0\leftrightarrow -W_{0\to 1}^{\rm rev}-(U_1-U_2)+W_{0\to 2}^{\rm rev}\ge 0.\label{eq.GB_Kelvin}
\end{eqnarray}
By using Eq.~(\ref{eq.GB_Kelvin}) to Eq.~(\ref{eq.GB_Snd}), we obtain the law of entropy non-decrease $S_2-S_1\ge 0$.
Meanwhile, it may be considered that the entropy in Eq.~(\ref{eq.GB_S}) is dependent on the choice of the reference heat reservoir through the temperature $T_o$.
In the GB approach, it is shown that Eq.~(\ref{eq.GB_S}) is in fact {\it independent} of $T_o$: 
The GB entropy is shown to be a state quantity {\it without} an explicit calculation of the reversible work, and any heat reservoir at any temperature can be used for the definition of entropy as the heat reservoir merely serves as an auxiliary system (see Sec.~7.4 of Ref.~\cite{GB2005} for their proof).
In contrast, in the present approach, we evaluated the reversible work explicitly by the use of the Carnot machine and showed that the entropy is independent of the choice of the reference heat reservoir (Eq.~(\ref{eq.S_Kelvin})). 

An advantage of the GB entropy Eq.~(\ref{eq.GB_S}) to the Clausius entropy Eq.~(\ref{eq.S_def}) is that the GB entropy is determined only from the measurement of $W_{0\to 1}^{\rm rev}$ for a reversible process with a heat reservoir at temperature $T_o$ and $W_{0\to 1}^{\rm ad}=U_0-U_1$ for an adiabatic process, with no need to measure heat.
But, these two expressions are related through the expression Eq.~(\ref{eq.S_Kelvin}). 
See Table~\ref{table} for the summary of these three expressions.

\begin{table}[t!]
\begin{tabular}{lll}\hline
Clausius (Eq.~(\ref{eq.S_def})) & Gyftopoulos and Beretta~\cite{GB2005} (Eq.~(\ref{eq.GB_S})) \ \ \ & This study (Eq.~(\ref{eq.S_Kelvin}))\\ \hline\hline
$\displaystyle S_1=S_0+\int_0^1 \frac{\delta Q}{T_R}$ & $\displaystyle S_1=S_0+\frac{W_{0\to 1}^{\rm rev}-(U_0-U_1)}{T_o}$ & $\displaystyle S_1=S_0+\int_0^1 \frac{\delta Q}{T}$\\
Changeable temperature $T_R$, heat \ \ \ & Fixed temperature $T_o$, work & Fixed temperature $T_o$, heat \\ \hline
\end{tabular}
\caption{Apparently different but equivalent three expressions of entropy: 
The Clausius expression uses heat divided by the changeable temperature of a heat reservoir; The Gyftopoulos-Beretta expression uses work divided by the fixed temperature of a single heat reservoir; The expression in this study uses heat divided by the system's temperature while a single heat bath at fixed temperature is involved.}\label{table}
\end{table}

\subsection{Clausius entropy using isothermal heat}\label{sec_isothermal}
By using the Clausius entropy Eq.~(\ref{eq.S_def}) given $S_0$ as the entropy of a reference state, we can calculate the entropy of another state along any reversible path. In particular, when we choose a reversible path that consists of isothermal processes with a single heat reservoir at temperature $T_o$ and adiabatic processes, we have the Clausius entropy in terms of isothermal heat:
\begin{eqnarray}
S_1=S_0+\frac{Q_{{\rm iso}, 1}}{T_o}, \ \ S_2=S_0+\frac{Q_{{\rm iso}, 2}}{T_o},\label{eq.S_iso}
\end{eqnarray}
where $Q_{{\rm iso}, 1}\equiv \int_0^1 \delta Q$ and $Q_{{\rm iso}, 2}\equiv \int_0^2 \delta Q$ for isothermal processes.
With this entropy in terms of the isothermal heat, it is possible to derive the law of entropy non-decrease Eq.~(\ref{eq.S_non}) directly from the Kelvin-Planck principle applied to the following cyclic process (see Sec.~8.11 of Ref.~\cite{ZD1997} for a similar argument):
Consider that the equilibrium state $A_1=A_1(T_1, V_1)$ is brought to another equilibrium state $A_2=(T_2, V_2)$ adiabatically in an arbitrary manner.
The equilibrium state $A_2$ is then brought to the equilibrium state $A_{2'}=A_{2'}(T_o,V_{2'})$ reversibly along the adiabatic curve.
By attaching a prepared heat reservoir at temperature $T_o$ to the system, the equilibrium state $A_{2'}$ is further brought to the equilibrium state $A_{1'}=A_{1'}(T_o,V_{1'})$ reversibly
along the isothermal curve until it intersects the adiabatic curve with the original equilibrium state $A_1=A_1(T_1,V_1)$ on it. Finally, after detaching the heat reservoir from the system,
the equilibrium state $A_{1'}$ is brought to the original equilibrium state $A_1=A_1(T_1,V_1)$ reversibly along the adiabatic curve.

The work extracted from the whole cyclic process $A_1 \stackrel{\rm ad}{\longrightarrow} A_2 \stackrel{\rm ad/rev}{\longrightarrow} A_{2'} \stackrel{\rm iso/rev}{\longrightarrow} A_{1'} \stackrel{\rm ad/rev}{\longrightarrow} A_1$ is given as
\begin{eqnarray}
W_{\rm cyc}&&=W_{1\to 2}^{\rm ad}+W_{2\to 2'}^{\rm ad/rev}+W_{2'\to 1'}^{\rm iso/rev}+W_{1'\to 1}^{\rm ad/rev}\nonumber\\
&&=U_1-U_2+U_2-U_{2'}+W_{2'\to 1'}^{\rm iso/rev}+U_{1'}-U_1\nonumber\\
&&=U_{1'}-U_{2'}+W_{2'\to 1'}^{\rm iso/rev}\nonumber\\
&&=U_{1'}-U_{2'}+U_{2'}-U_{1'}-T_o(S_{2'}-S_{1'})\nonumber\\
&&=T_o(S_{1'}-S_{2'}),\label{eq.Kelvin_iso}
\end{eqnarray}
where $W_{2'\to 1'}^{\rm iso/rev}=U_{2'}-U_{1'}-T_o(S_{2'}-S_{1'})$, which stands for the work extracted from the reversible isothermal process, coincides with the minus of the Helmholtz free energy difference of the system before and after the isothermal process.
By applying the Kelvin-Planck principle Eq.~(\ref{eq.Kelvin}) to Eq.~(\ref{eq.Kelvin_iso}), we have $W_{\rm cyc}=T_o(S_{1'}-S_{2'})=T_o(S_1-S_2)\le 0$ and thus we derive the law of entropy non-decrease $S_1\le S_2$.

The outline of the above derivation is essentially the same as those in Secs.~\ref{sec_simple} and \ref{sec_compound}.
An important difference is that the calculation of the work here is done by using the Helmholtz free energy difference for a reversible isothermal process, while it is done by using the reversible work Eq.~(\ref{eq.W_gamma_S}) for a reversible temperature-changing process with a Carnot machine in Secs.~\ref{sec_simple} and \ref{sec_compound}. 
With this process using a Carnot machine, we were able to apply the Kelvin-Planck principle to show the non-decrease of the entropy Eq.~(\ref{eq.S_Kelvin}), which is essentially the same entropy as the Clausius entropy Eq.~(\ref{eq.S_def}) in terms of non-isothermal heat.

\subsection{Characterization of entropy as extra thermodynamic cost to create nonuniformity}
\begin{figure}[t!]
\centering
\includegraphics[scale=0.55]{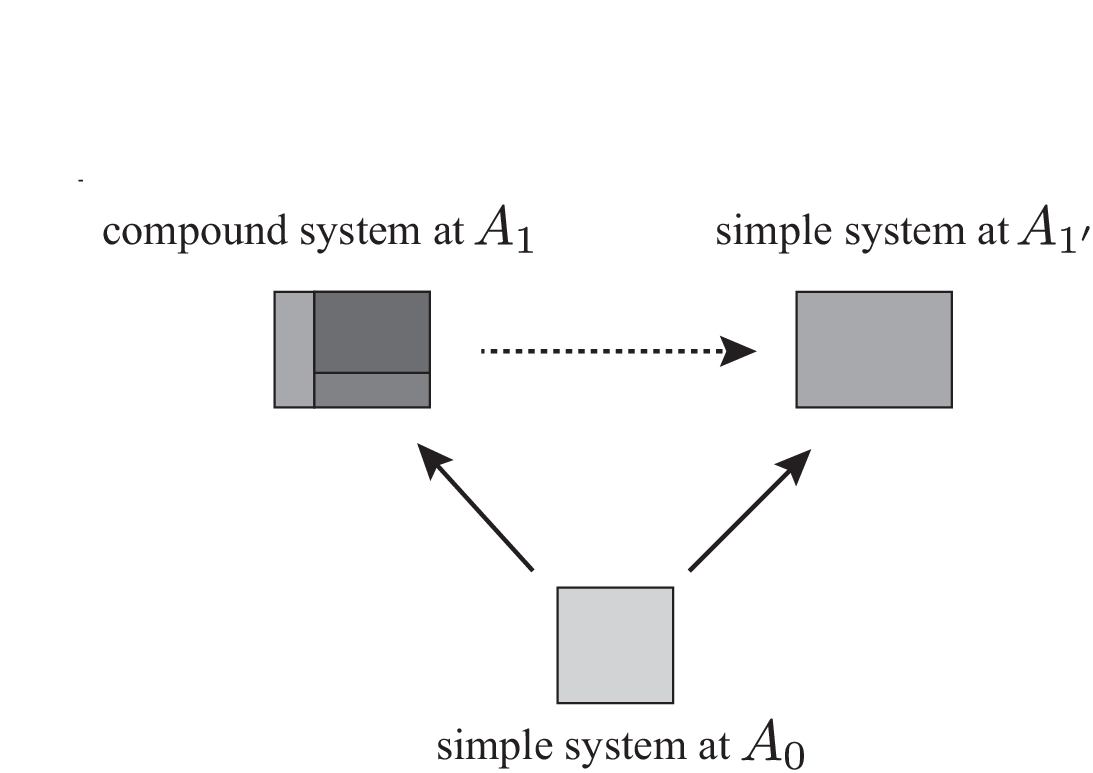}
\caption{A compound system at state $A_1$ and a simple system at state $A_{1'}$ of the same internal energy and volume are created from a simple system at state $A_0$. An adiabatic process $A_1 \stackrel{\rm ad}{\longrightarrow} A_{1'}$, which occurs spontaneously after the removal of the insulated walls separating the subsystems of the compound system, is realized only when Eq.~(\ref{eq.W_he}) is satisfied.}
\label{fig_simple_compound}
\end{figure}

In light of the discussion so far, we may characterize entropy as an extra thermodynamic cost to create nonuniformity in the system.
To this end, let us consider a setup in Fig.~\ref{fig_simple_compound}: From a simple system at state $A_0$, a compound system at state $A_1$ constituted with $M$ subsystems and a simple system at state $A_{1'}$ are created, where the internal energies and volumes of the states $A_1$ and $A_{1'}$ are set to be the same.
The thermodynamic cost, the work taken for creating each state by combining reversible adiabatic processes and reversible temperature-changing processes with a Carnot machine (Fig.~\ref{fig_entropy}), is calculated as the minus of the extracted work as
\begin{eqnarray}
&&-W_{0\to 1}^{\rm rev}=-(U_0-U_1-T_o(S_0-S_1)),\label{eq.W_01_com}\\
&&-W_{0\to 1'}^{\rm rev}=-(U_0-U_{1'}-T_o(S_0-S_{1'})),\label{eq.W_01dash_sim}
\end{eqnarray}
where $U_1=\sum_{j=1}^M U_{1,j}$ and $S_1=\sum_{j=1}^M S_{1,j}$, and $U_1=U_1'$.
Consider a cyclic process $A_0 \stackrel{\rm rev}{\longrightarrow} A_1 \stackrel{\rm ad}{\longrightarrow} A_{1'} \stackrel{\rm rev}{\longrightarrow} A_0$, where during $ A_1 \stackrel{\rm ad}{\longrightarrow} A_{1'}$ the allowed adiabatic operation is to remove all the insulated walls separating the subsystems so that the compound system at state $A_1$ {\it spontaneously} changes to the simple system at state $A_{1'}$; the work taken for this process is assumed to vanish as $-W_{1\to 1'}^{\rm ad}=0$.
The application of Kelvin-Planck principle to this cyclic process yields $W_{0\to 1}^{\rm rev}\le W_{0\to 1'}^{\rm rev}$, which can be expressed in terms of the work cost as
\begin{eqnarray}
-W_{0\to 1}^{\rm rev}\ge -W_{0\to 1'}^{\rm rev}.\label{eq.W_he}
\end{eqnarray}
Note that this is equivalent to the law of entropy non-decrease $S_1\le S_{1'}$ as confirmed from Eqs.~(\ref{eq.W_01_com}) and (\ref{eq.W_01dash_sim}) with $U_1=U_{1'}$.
The expression Eq.~(\ref{eq.W_he}) implies that the spontaneous process from $A_1$ to $A_{1'}$ occurs only when $A_1$ is more {\it costly} than $A_{1'}$ to create from $A_0$ thermodynamically; the nonuniform, low-entropy (ordered) structure is hard to create, which is easily seen by comparing Eqs.~(\ref{eq.W_01_com}) and (\ref{eq.W_01dash_sim}) with $U_1=U_{1'}$. This may give a characterization of entropy as an extra thermodynamic cost that needs to be paid to create nonuniformity in the system, which makes it possible to distinguish between two macroscopic systems of the same internal energy and volume.

As an example, we consider entropy of an ideal gas. We create states $A_1(\{T_{1,1},V/2\},\{T_{1,2},V/2\})$ ($M=2$) and $A_{1'}(T_{1'},V)$ from a reference state $A_0(T_0, V)$. The internal energy $U_1=U_{1,1}+U_{1,2}$ and $U_{1'}$ coincide as no thermodynamic cost is assumed to be necessary to remove the insulated wall partitioning the two subsystems. 
From Eq.~(\ref{eq.S_Kelvin}), the entropy change from $A_0$ to $A_1$ reads
\begin{eqnarray}
S_1-S_0=\int_0^1 \frac{\delta Q}{T}=\int_{T_0}^{T_{1,1}}\frac{C_V}{2T}dT+\int_{T_0}^{T_{1,2}}\frac{C_V}{2T}dT=C_V\ln \frac{\sqrt{T_{1,1}T_{1,2}}}{T_0},
\end{eqnarray}
where $C_V$ denotes the constant-volume heat capacity of the ideal gas. In the same manner, we obtain the entropy change from $A_0$ to $A_{1'}$ as
\begin{eqnarray}
S_{1'}-S_0=\int_0^{1'} \frac{\delta Q}{T}=\int_{T_0}^{T_{1'}}\frac{C_V}{T}dT=C_V\ln \frac{T_{1'}}{T_0}=C_V\ln \frac{T_{1,1}+T_{1,2}}{2T_0},
\end{eqnarray}
where we used $T_{1'}=(T_{1,1}+T_{1,2})/2$ obtained from $U_1=U_{1,1}+U_{1,2}=U_{1'}$ in the last equality. From Eqs.~(\ref{eq.W_01_com}) and (\ref{eq.W_01dash_sim}), the works taken for these changes read
\begin{eqnarray}
&&-W_{0\to 1}^{\rm rev}=-(U_0-U_1)-C_VT_o\ln \frac{\sqrt{T_{1,1}T_{1,2}}}{T_0},\label{eq.W_01_com_ig}\\
&&-W_{0\to 1'}^{\rm rev}=-(U_0-U_{1'})-C_VT_o\ln \frac{T_{1,1}+T_{1,2}}{2T_0},\label{eq.W_01dash_sim_ig}
\end{eqnarray}
respectively, where we note $U_1=U_{1'}$. Because of $(x+y)/2 \ge \sqrt{xy}$ for $x, y \ge 0$, we have $-W_{0\to 1}^{\rm rev}\ge -W_{0\to 1'}^{\rm rev}$.

\section{Conclusion}
We derived the law of entropy non-decrease Eq.~(\ref{eq.S_non}) from the Kelvin-Planck principle Eq.~(\ref{eq.Kelvin}) for both simple and compound systems without using the Clausius inequality Eq.~(\ref{eq.Clausius_ineq}), where the process that utilizes the Carnot machine operating between the system and the heat reservoir played the essential role in the derivation.
Our entropy Eq.~(\ref{eq.S_Kelvin}) is equivalent to the Clausius entropy Eq.~(\ref{eq.S_def}) and the Gyftopoulos-Beretta entropy Eq.~(\ref{eq.GB_S}), establishing a thermodynamic consistency. We also characterized entropy as the extra thermodynamic cost that needs to be paid to create nonuniformity in the system.
We hope that the present derivation elucidates the principle of thermodynamics.
It is also of interest to apply the present formulation of entropy to nonequilibrium entropies that determine adiabatic accessibility between nonequilibrium states~\cite{LY2013}.

\begin{acknowledgment}


The author is grateful to A. Calvo Hern\'{a}ndez for valuable comments on the earlier version of the manuscript.
This work was supported by JSPS KAKENHI Grant Number 22K03450.

\end{acknowledgment}

\appendix
\section{Proof that Eq.~(\ref{eq.W_gamma_S}) serves as the maximum work}\label{sec_appendix}

We show that Eq.~(\ref{eq.W_gamma_S}) serves as the maximum work extracted from a system changing from one equilibrium state $A_0$ to another equilibrium state $A_1$ with the aid of a heat reservoir at fixed temperature $T_o$. 
The argument is based on Sec.~3(b) of Ref.~\cite{LY2013}.

Consider a thermodynamic process where a system changes its state from one equilibrium state $A_0$ to another $A_1$. 
There is no need for the system's temperature to match the fixed temperature $T_o$ of the heat reservoir during the process including the initial and final states $A_0$ and $A_1$.
The entropy change of the total system, i.e., the system and the heat reservoir~\cite{note3}, reads 
\begin{eqnarray}
S_1-S_0+\Delta S_R \ge 0,\label{eq.DS_total}
\end{eqnarray}
where $\Delta S_R$ denotes the entropy change of the heat reservoir. The minus of the internal energy change of the total system, which is equivalent to the work $W$ extracted from the system, reads 
\begin{eqnarray}
W=U_0-U_1-\Delta U_R=U_0-U_1-T_o \Delta S_R,\label{eq.W_total}
\end{eqnarray}
where $\Delta U_R=T_o\Delta S_R$ denotes the internal energy change of the heat reservoir. By applying the inequality Eq.~(\ref{eq.DS_total}) to Eq.~(\ref{eq.W_total}), we obtain
\begin{eqnarray}
W\le U_0-U_1-T_o(S_0-S_1).\label{eq.W_ineq}
\end{eqnarray}
In the case using a reversible isothermal process, the maximum work, the right-hand side of Eq.~(\ref{eq.W_ineq}), is nothing but the Helmholtz free energy difference (Sec.~\ref{sec_isothermal}). But this is not the only case, and as we see in Eq.~(\ref{eq.W_gamma_S}), the equality is non-trivially attained also for the reversible temperature-changing process with a Carnot machine.


\begin{thebibliography}{9}
\bibitem{note1} We assume that a working substance of the Carnot machine returns to the initial state after the operation.
\bibitem{note3} We may use a Carnot machine to change the system's state during this thermodynamic process. As we assume that a working substance of a Carnot machine returns to the initial state after the operation, the entropy change of the working substance vanishes and does not contribute to the left-hand side of Eq.~(\ref{eq.DS_total}).
\end{thebibliography}
\end{document}